\documentclass[12pt]{article}
  
\usepackage{calrsfs,bbm,bm}
\usepackage{axodraw,color,dsfont,cancel,graphicx,mathrsfs,yfonts}
\usepackage{amsmath}
\usepackage{amssymb}
 
\usepackage{exscale}

\usepackage{hyperref}

\font\fourteenbf=cmbx12 scaled\magstep1

\def\a0size{6}

\newcommand{\lsi}{\raise0.3ex\hbox{$<$\kern-0.75em\raise-1.1ex\hbox{$\sim$}}}
\newcommand{\gsi}{\raise0.3ex\hbox{$>$\kern-0.75em\raise-1.1ex\hbox{$\sim$}}}

\renewcommand{\vec}[1]{{\bm #1}}

\newcommand{\be}{\begin{equation}}
\newcommand{\ee}{\end{equation}}
\newcommand{\baed}{\begin{aligned}}
\newcommand{\eaed}{\end{aligned}}

\begin{document} 

\setlength{\baselineskip}{0.6cm}
\newcommand{\figysize}{16.0cm}
\newcommand{\figtopspace}{\vspace*{-1.5cm}}
\newcommand{\figbottomspace}{\vspace*{-5.0cm}}
  

\begin{titlepage}
\begin{flushright}
BI-TP 2012/05
\\
\end{flushright}
\begin{centering}
\vfill

{\fourteenbf 
\centerline{ 
Thermal production of ultrarelativistic right-handed neutrinos:       }
\centerline{ 
Complete leading-order results      }
}

\vspace{1cm}

Denis Besak,  \footnote{denis.besak@gmx.de}
Dietrich B\"odeker  \footnote{bodeker@physik.uni-bielefeld.de}

\vspace{.6cm} { \em 
Fakult\"at f\"ur Physik, Universit\"at Bielefeld, D-33615 Bielefeld, Germany
}

\vspace{2cm}
 
{\bf Abstract}

\end{centering}
 
\vspace{0.5cm}
\noindent
The thermal production of relativistic right-handed Majorana neutrinos is of
importance for models of thermal leptogenesis in the early
Universe. Right-handed neutrinos can be produced both by $1 \leftrightarrow 2$
decay or inverse decay and by $2 \to 2$ scattering processes. In a previous
publication we have studied the production via $ 1 \leftrightarrow 2 $
(inverse) decay processes.  There we have shown that multiple scattering
mediated by soft gauge boson exchange also contributes to the production rate
at leading order and gives a strong enhancement.  Here we complete the
leading order calculation by adding $2 \to 2$ scattering processes involving
either electroweak gauge bosons or third-generation quarks.  We find that
processes with gauge interactions give the most important contributions. We
also obtain a new sum rule for the Hard Thermal Loop resummed fermion
propagator.

\vspace{0.5cm}\noindent

 
\vspace{0.3cm}\noindent
 
\vfill \vfill
\noindent
 
\end{titlepage}

\tableofcontents
 
\section{Introduction and motivation}
\label{sc:intro} 

One of the outstanding problems of standard cosmology is to explain the origin
of the asymmetry between matter and antimatter. Without such an asymmetry, all
the structures we observe today would have never formed and mankind would not
exist. The asymmetry can be expressed as the \emph{baryon-to-photon ratio} 

\begin{align} 
 \frac{n_B}{n_\gamma} = (6.19 \pm 0.15 )\cdot 10^{-10} 
\end{align} 
whose numerical value is obtained from a combined analysis of data for
large-scale structure and the spectrum of the Cosmic Microwave
Background~\cite{wmap7}. 

In order to obtain a net baryon asymmetry, only three conditions need
to be met, as outlined by Sakharov in his seminal
paper~\cite{Sakharov}. Yet, providing a model that can successfully
explain the measured baryon-to-photon ratio remains a challenging
task.  Several different scenarios how to realize the Sakharov
conditions have been devised~\cite{Baryogenesis}.  In the last decade,
leptogenesis~\cite{Fukugita} has become very popular. The basic idea
of most leptogenesis models is to extend the 
Standard Model (SM) by adding heavy right-handed neutrinos $ N _
{\rm R } $ with Majorana mass $ M _ N $. In the simplest realization they interact with the SM
particles via a Yukawa coupling to ordinary, left-handed leptons $
\ell _ {\rm L } $ and the Higgs bosons $ \varphi $ as follows:
\begin{align}\label{Lint}
   \mathcal{L}_{\rm int} = h_{ij} \overline{N _ {\rm R } }_i {\widetilde \varphi}^\dagger 
   \ell_{{\rm L } j}
   + \mbox{ h.c. }
   .
\end{align}
Here $\widetilde \varphi \equiv i \sigma ^ 2 \varphi
^ \ast $ with the Pauli matrix $ \sigma ^ 2 $ is the isospin conjugate
of $ \varphi $.  Furthermore, the indices $i,j$ label the
fermion families, and $h_{ij}$ is the Yukawa coupling matrix which need not be
diagonal.

The Majorana neutrinos are unstable and decay both into leptons and
antileptons, $N \to \ell \varphi $, $ N \to \bar {\ell} \overline{ \varphi }
$. The $  CP $ 
symmetry is violated and the corresponding decay rates are not
equal. Therefore, an excess of antileptons over leptons can be generated. The
resulting asymmetry is converted into an excess of baryons over antibaryons
via the sphaleron transitions which conserve $B - L$ but violate $B +
L$~\cite{kuzmin}. In addition to providing a source for the measured
baryon asymmetry, this scenario offers a framework to explain the smallness of
the neutrino masses via the \emph{seesaw mechanism}~\cite{seesaw}. This
twofold virtue is what makes the scenario of leptogenesis particularly
appealing.

So far there is no complete leading order treatment of leptogenesis, for the
case of relativistic Majorana neutrinos. It has been argued that there are
theoretical uncertainties of a factor 2 or more \cite{interference}. The
purpose of this paper is to make a step towards a complete leading order
treatment. We compute the complete leading order production rate of Majorana
neutrinos in the symmetric phase of the electroweak theory, \footnote{Results
  for the broken phase can be found in Ref.~\cite{shaposhnikov}. The
  production rate in the low temperature regime has recently been calculated
  at next-to-leading order \cite{laine-nlo} correcting an earlier calculation
  in Ref.~\cite{salvio}.  }  in the regime $ T \gg M _ N $. It is one
ingredient in the set of kinetic equations which describe leptogenesis. At the
end of inflation the number density of right-handed neutrinos should be
negligible, and therefore the production rate determines the initial
conditions for leptogenesis. Thus it is particularly important in the
so-called weak washout regime, where the Majorana neutrinos are not close to
equilibrium.  It is essential when their number density always remains far
below the equilibrium density. This is the case in scenarios where the
asymmetry is generated via oscillations between the right-handed neutrinos
\cite{akhmedov,asaka-bau}.  These types of models also can potentially explain
both the baryon asymmetry and dark matter \cite{asaka-bau,asaka-dm}.

Two types of processes contribute to the leading order production rate. The
first includes the inverse decay $ \varphi \ell \to N $. At high temperature
all masses are smaller than the typical energies. Then the momenta in this
process are nearly collinear. It is therefore quite sensitive to thermal
mass effects~\cite{giudice,kiessig,anisimov}. The above process is
kinematically forbidden at high temperature, and the decay $ \varphi \to \ell
N $ becomes possible.~\footnote{In \cite{giudice} a lepton mass corresponding
  to soft fermionic excitations was used. However, at leading order the
  dominant contribution is due to hard 
  leptons, and one has to use the so-called 
asymptotic mass instead \cite{anisimov}. This was improved 
  in Ref.~\cite{kiessig} by resumming the Hard
  Thermal Loop fermion self-energy. For hard momenta this is equivalent, at
  leading order, to using the asymptotic mass.}  In \cite{anisimov} it was
shown that processes with additional scattering mediated by soft electroweak
gauge bosons exchange also contribute at leading order.~\footnote{Among other
  contributions, they include a resummation of a finite width of the Standard
  Model particles, the role of which was recently emphasized in
  \cite{quantum}.}  The soft scattering opens new decay and inverse decay
channels.  One has to sum over arbitrary numbers of such interactions which
was done in Ref.~\cite{anisimov}, where it turned out that including them
increases the rate by almost a factor 3.

However, the calculation in \cite{anisimov} is still incomplete: $2\to
2$ scattering processes involving hard ($ p \sim T $) electroweak gauge bosons or
third-generation quarks contribute to the leading order production
rate as well. Quark-initiated processes have been taken into account
by many authors (see e.g. \cite{luty,pilaftsis,hahn,buchmuller-cmb}),
explicit results for the corresponding rate were shown in
\cite{pilaftsis,buchmuller-cmb}. Gauge boson scattering processes have
been considered in Ref.~\cite{giudice,pilaftsis}. They contain a 
leading order contribution with the exchange of a soft fermion in the $ t
$-channel,
 which is infrared divergent
when medium effects are not taken into account. 
So far no consistent
computation, which properly treats the soft exchange contribution has
been carried out.  The goal of this publication is to close this gap and
complete the treatment of the thermal production of Majorana neutrinos
in a hot electroweak plasma.

This paper is organized as follows. In Sec.~\ref{sc:rate} we review the 
processes which can create a population of right-handed
neutrinos, and present equations for the differential production rate.
The results of
Ref.~\cite{anisimov} for the collinear emission processes 
including
soft gauge boson exchange scattering are 
briefly reviewed in Sec.~\ref{sc:coll}. Then in Sec.~\ref{sc:2by2} we
describe the calculation of the production rate due to $2\rightarrow 2$ scattering
processes off hard quarks or electroweak gauge bosons. Our main result,  the total
$ 2\to 2 $ rate, is shown in  Eq.~(\ref{222formel}) of
Sec.~\ref{sc:results}. This
section also contains numerical results for the total and differential rates.
In
Sec.~\ref{sc:conclusions} we then conclude.  Appendix
\ref{ap:details} contains technical details of the
calculation of the  $2\to 2$ scattering rate with
hard momentum transfer.  In Appendix \ref{ap:sumrule} we show that the
Hard Thermal Loop resummed propagator satisfies a sum rule which
greatly simplifies the calculation of the $ 2 \to 2 $ rate with soft fermion exchange in
Sec.~\ref{sc:2by2}.

\noindent {\bf  Conventions and notation} The signature of the metric is $ + $ $ - $ $ -
$ $ - $, 4-vectors are denoted by lower case italics, 3-vectors by boldface
italics.

\section{Production rate and processes}      
\label{sc:rate} 

We consider a hot electroweak plasma that is fully equilibrated,
except for the right-handed Majorana neutrinos, which are assumed to
have negligible number density.  We compute their production rate $
\Gamma $, i.e., the number of right-handed neutrinos which are
thermally produced per unit time and unit volume, at leading order in
the Yukawa interaction (\ref{Lint}).  We focus on the lightest
Majorana neutrinos $N_1 \equiv N$, which we assume to be the dominant
source of lepton asymmetry. We include all contributions of leading
order in the SU(2) and U(1) gauge couplings $g$ and $g'$, the top
quark Yukawa coupling constant $h _ t$ and the Higgs self-coupling
$\lambda$.  For the power counting we assume that all these couplings
are of the same order and collectively refer to them by $g$.  All
other SM couplings are neglected.  We perform the computation in the
high-temperature regime where $M_N \ll T$. This allows us to formally
treat the mass of the Majorana neutrino as being soft, $ M_N \sim g T
$, and therefore parametrically of the same order as the thermal Higgs
and lepton masses. The leading order rate is then of order $ h ^ 2 g ^ 2 $
where
\begin{align} 
  h^2 \equiv \sum_j |h_{1j}|^2
  \, 
  .
\end{align} 

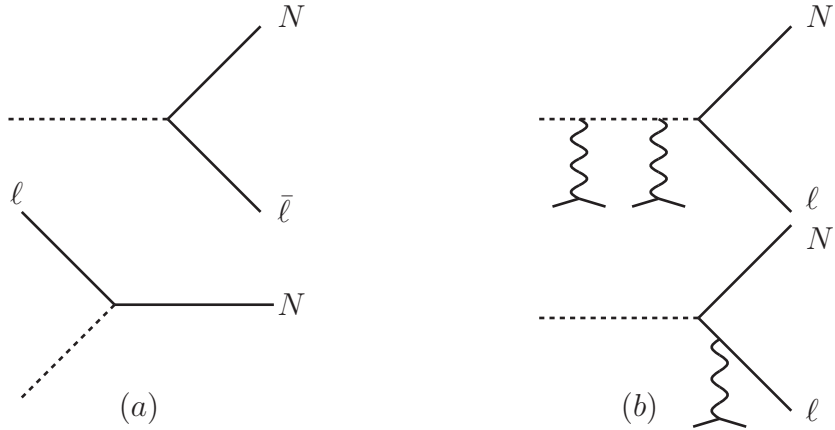
\begin{figure}[!t] 
\centering
\fcolorbox{white}{white}{
  \begin{picture}(0,150) (200,-70)
    \SetWidth{1.0}
  \DashLine(80,50)(140,50){2}
  \Line(140,50)(175,85)
  \Line(140,50)(175,15)
  \Text(182,85)[lb]{\normalsize{$N$}}
  \Text(182,11)[lb]{\normalsize{$\bar \ell$}}
  \Line(120,-20)(180,-20)
  \DashLine(85,-55)(120,-20){2}
  \Line(120,-20)(85,15)
  \Text(182,-24)[lb]{\normalsize{$N$}}
  \Text(81,18)[lb]{\normalsize{$\ell$}}
  \Text(122,-65)[lb]{\normalsize{$(a)$}}
  \DashLine(280,50)(340,50){2}
  \Line(340,50)(375,85)
  \Line(340,50)(375,15)
  \Photon(295,50)(295,20){3}{3}
     \Line(285,17)(295,20)
     \Line(305,17)(295,20)
  \Photon(325,50)(325,20){3}{3}
     \Line(315,17)(325,20)
     \Line(335,17)(325,20)
  \Text(382,85)[lb]{\normalsize{$N$}}
  \Text(382,16)[lb]{\normalsize{$\ell$}}
  \DashLine(280,-25)(340,-25){2}
  \Line(340,-25)(375,10)
  \Line(340,-25)(375,-60)
  \Photon(348,-63)(348,-33){3}{2.9}
     \Line(338,-66)(348,-63)
     \Line(348,-63)(358,-66)
  \Text(382,1)[lb]{\normalsize{$N$}}
  \Text(382,-64)[lb]{\normalsize{$\ell$}}
  \Text(312,-65)[lb]{\normalsize{$(b)$}}

\end{picture}
}
\caption{Leading order contributions to the production of right-handed 
  neutrinos~$ N $. ($ a $) Decay and
  recombination processes which
  occur when the asymptotic thermal masses are such that  
  they are kinematically allowed. ($ b $) Examples for
  processes, where exchanges of soft gauge bosons with particles in
  the plasma have been added.  
  Higgs bosons are denoted with a dashed line and gauge bosons with a
  wiggled line. Fermions are represented by solid lines.} 

\label{fg:12}
\end{figure}

There are two types of processes by which right-handed neutrinos are produced
at leading order. Among the first are the $1\leftrightarrow 2$ processes
(see  Fig.~\ref{fg:12}($ a $))
\begin{align} 
  \varphi  \ell \to N 
  ,  \quad 
  \varphi  \to \bar \ell N
  , \quad 
  \ell \to \overline \varphi  N
   \label{decay}
   .
\end{align}
At low temperature only the first one is kinematically
allowed. When $ T $ increases,  the thermal masses $ m _
\varphi $ and $ m _ \ell $ (see Eqs.~(\ref{mphi}), (\ref{mell}))
increase. On the other hand, the thermal mass of the right-handed neutrinos is
negligible.  Therefore $ m _ \varphi  + m _ \ell $ can 
become larger than $ M _ N $, so that the inverse $ N $-decay is kinematically
no longer
allowed. At even higher temperature $m_\varphi > M_N + m_\ell $. Then the
Higgs decay in (\ref{decay}) becomes possible. The third process in 
Eq.~(\ref{decay}) is not allowed at any temperature because $m_\varphi > m_\ell$. 
However, including multiple interactions with soft gauge boson
exchange as shown in Fig.~\ref{fg:12}($ b $) changes this picture. Then  the
decay/recombination processes do occur at any temperature and it also
makes the process $\ell \to \overline{ \varphi } N$ possible.  As was
demonstrated in \cite{anisimov}, these processes with an arbitrary number of 
soft gauge interactions contribute at leading
order.

\begin{figure}[!t] 
\centering
  \begin{picture}(0,235) (220,-80) 
    \SetWidth{1.0}

    \Line(70,100)(95,125)
    \Line(70,150)(95,125)
    \DashLine(95,125)(135,125){2}
    \Line(135,125)(160,150)
    \Line(135,125)(160,100)
    \Text(62,95)[lb]{\normalsize{$\bar t$}}
    \Text(162,152)[lb]{\normalsize{$N$}}
    \Text(162,96)[lb]{\normalsize{$\ell$}}
    \Text(62,152)[lb]{\normalsize{$Q_3$}}
    \Text(107,92)[lb]{\normalsize{$(a)$}}

    \Line(230,108)(310,108)
    \Line(230,148)(310,148)
    \DashLine(270,108)(270,148){2}
    \Text(225,145)[lb]{\normalsize{$\bar t$}}
    \Text(312,104)[lb]{\normalsize{$N$}}
    \Text(225,104)[lb]{\normalsize{$\bar \ell $}}
    \Text(314,144)[lb]{\normalsize{$\bar Q_3$}}
    \Text(267,92)[lb]{\normalsize{$(b)$}}

    \Line(370,108)(450,108)
    \Line(370,148)(450,148)
    \DashLine(410,108)(410,148){2}
    \Text(357,145)[lb]{\normalsize{${Q}_3$}}
    \Text(452,104)[lb]{\normalsize{$N$}}
    \Text(365,104)[lb]{\normalsize{$\bar \ell  $}}
    \Text(454,144)[lb]{\normalsize{$t$}}
    \Text(407,92)[lb]{\normalsize{$(c)$}}

    \DashLine(10,60)(50,60){2}
    \Line(50,60)(50,20)
    \Photon(50,20)(90,20){3}{4}
    \Line(50,60)(90,60)
    \Line(10,20)(50,20)
    \Text(95,58)[lb]{\normalsize{$N$}}
    \Text(5,18)[lb]{\normalsize{$\ell$}}
    \Text(54,39)[lb]{\normalsize{$\ell$}}
    \Text(107,1)[lb]{\normalsize{$(d)$}}
    \DashLine(130,60)(170,60){2}
    \Line(130,20)(210,20)
    \DashLine(170,60)(170,20){2}
    \Photon(170,60)(210,60){3}{4}
    \Text(212,18)[lb]{\normalsize{$N$}}
    \Text(124,18)[lb]{\normalsize{$\ell$}}

    \Line(245,15)(270,40)
    \Line(270,40)(310,40)
    \DashLine(310,40)(335,15){2}
    \Line(310,40)(335,65)
    \Photon(245,65)(270,40){3}{4}
    \Text(338,66)[lb]{\normalsize{$N$}}
    \Text(240,9)[lb]{\normalsize{$\ell$}}
    \Text(290,43)[lb]{\normalsize{$\ell$}}
    \Text(347,1)[lb]{\normalsize{$(e)$}}
    \Line(365,60)(445,60)
    \DashLine(405,60)(405,20){2}
    \DashLine(405,20)(445,20){2}
    \Photon(365,20)(405,20){3}{4}
    \Text(449,58)[lb]{\normalsize{$N$}}
    \Text(360,58)[lb]{\normalsize{$\ell$}}

    \Line(195,-45)(220,-70)
    \Line(195,-45)(220,-20)
    \DashLine(155,-45)(195,-45){2}
    \DashLine(155,-45)(130,-20){2}
    \Photon(155,-45)(130,-70){3}{4}    
    \Text(223,-18)[lb]{\normalsize{$N$}}
    \Text(225,-75)[lb]{\normalsize{$\ell$}}
    \Line(295,-25)(295,-65)
    \Line(295,-65)(335,-65)
    \Line(295,-25)(335,-25)
    \DashLine(255,-25)(295,-25){2}
    \Photon(255,-65)(295,-65){3}{4}
    \Text(340,-27)[lb]{\normalsize{$N$}}
    \Text(340,-68)[lb]{\normalsize{$\ell$}}
    \Text(299,-49)[lb]{\normalsize{$\ell$}}
    \Text(237,-87)[lb]{\normalsize{$(f)$}} 
  \end{picture}
\caption{$ 2 \to 2 $ scattering contributions to Majorana neutrino production.}
\label{fg:2by2}
\end{figure}
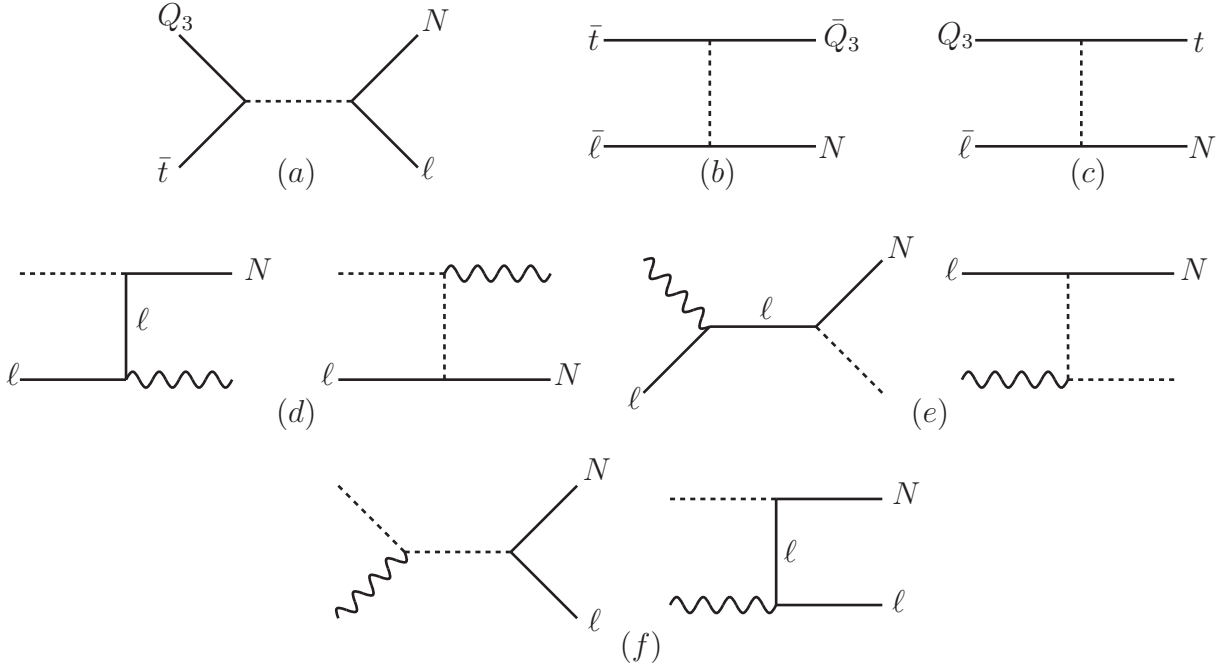

The second type are the $ 2 \to 2 $ hard particle scattering processes shown in 
Fig.~\ref{fg:2by2}. They can also occur
at any temperature. There are processes involving quarks,
\begin{align} 
  \label{scatQ} 
  Q_3  \overline{ t  } \to \ell N, \quad t \ell \to Q_3 N
  ,\quad 
  \quad  \overline{ Q } _ { 3 } \ell \to \overline{ t }   N, 
\end{align} 
where $t$ denotes the right-handed
top quark and $Q_3$ the doublet of left-handed third-generation quarks. They
contribute at the order $h^2 h_t ^2 $  and thus need to be taken
into account in a complete leading order computation. Additionally, there are
processes involving SU(2) or U(1) gauge bosons $ V $, 
\begin{align} 
  \label{scatG} \ell 
\varphi \to N V, \quad \ell V \to N \overline{  \varphi } , \quad
\overline{ \varphi } V \to N \ell , 
\end{align} 
which contribute at order $h^2 g^2 $ or $h^2 g'^2$, which means that they are
also part of the leading order production rate. With each of these processes,
we also need to include their $ CP $ conjugate, where every particle is
replaced by its antiparticle. At leading order there is no $ CP $ violation,
and the rates are the same.

For thermal field theory calculations it is convenient to express the
rate $ \Gamma $ in terms of the self-energy of the right-handed
neutrinos.  One can define the self-energy $ \Sigma $ as usual if one
promotes the right-handed spinor $ N _ {\rm R } $ to a 4-component
Majorana-spinor $ N \equiv N _ {\rm R } + N _ {\rm R } ^ c $ where $ c
$ stands for charge conjugation.  Then $ \Gamma $ can be obtained from
the imaginary part of the retarded self-energy $\Sigma_{\rm ret} (k)=
\Sigma (k ^ 0 + i0^+,\vec k)$, where $ k ^ 0 $ and $ \vec k $ are the
(real) energy and the 3-momentum of the produced neutrinos, through
the relation~\cite{anisimov}
\begin{align}
   \label{rate} 
   (2\pi)^3 2 k ^ 0 
   \frac{d\Gamma }{d^3 k} =
   2 f _ { \rm F } (k ^ 0) 
     \operatorname {Im} \operatorname{Tr} \left[
     \cancel{k} \Sigma_{\rm ret} (k) \right]
   .
\end{align}
Here
$ f _ { \rm   F } $ is the Fermi-Dirac distribution function. 
Due to the Majorana-nature of $ N $ there are two types of diagrams
contributing to the self-energy which differ by the orientation of the
internal lines. Both types of diagrams give the same contribution.  We
consider only one of them, and multiply by 2 to obtain the correct rate.  This
amounts to using 
\begin{align} 
   \label{rateRH} 
   (2\pi)^3  2 k ^ 0 \frac{d\Gamma}{d^3 k} =
   4 f _ { \rm F } ( k ^ 0 ) \operatorname {Im}   \operatorname{Tr} \left[
       \cancel{k} P _ {\rm L }  \Sigma_{\rm ret} (k) \right]
\end{align}
with the left-handed projector $ P _ {\rm L } = (1 - \gamma  _ 5 ) /2 $. 

\section{
$ 1 \leftrightarrow 2 $ 
 scattering including soft gauge interactions}      
\label{sc:coll} 

\subsection{Kinematics} 
\label{sc:kinematics}

First consider the processes shown in Fig.~\ref{fg:12}.  All external momenta
in Fig.~\ref{fg:12}($ a $) are hard ($ p \sim T $). 
The mass of $ N $ as well as the thermal masses are soft, i.e., of order
$ g T $. Therefore the typical angles between the hard momenta are of order
$ g $. We assume $ g $ to be small, which implies that all momenta are nearly
collinear. Fig.~\ref{fg:12}($ b $) shows processes with additional interactions
mediated by soft electroweak gauge bosons. With these
soft interactions the momenta of $ \varphi $, $ \ell $ and $ N $ are still
nearly collinear.  Furthermore, the hard 4-momenta are close to the
light-cone, $ p ^ 2 \sim g ^ 2 T ^ 2 $. We denote by $ \vec v \equiv \vec k /|
\vec k | $ the unit 3-vector $ \vec v $ in the direction of the momentum $
\vec k $ of the right-handed neutrino.  For
all vectors the components parallel to $ \vec v $ are denoted by
\begin{align} 
    p _ \| \equiv \vec p \cdot  \vec v
    \label{pparallel} .
\end{align} 
We further define the light-like vector  $ v \equiv 
( 1,  {\vec v }  ) $. 
Then one has to account for three distinct momentum scales:
\begin{enumerate}
\item The emitting  particles 
  and the emitted particle have 
  $ p _ \| \sim T$.
\item All 3-momenta perpendicular to $ \vec v $ are soft, $ \vec p _\perp \sim
  g T $. Also all components  of the exchanged gauge boson momentum $ q $ 
  are soft, $ q \sim g T
  $.
\item Finally, all 4-momenta $ p $ satisfy
  $ v \cdot p = ( p  _ 0 - p _ \| ) \sim g ^ 2 T $.  
\end{enumerate}
For  momenta with $ p ^ 2 \sim g ^ 2 T ^ 2 $ one cannot neglect the modification of the
dispersion relations which is caused by the  interactions with the hot
plasma. For hard particles the dispersion relation  can be written as $
p ^ 2 = m ^ 2 $, where $ m $ is the so-called asymptotic mass
\cite{weldon-fermion}. \footnote{It is oftentimes referred to as $
  m_\infty $.}  For the Higgs and the lepton doublets they are given
by \footnote{For scalars the asymptotic mass is the same as the
  thermal self-energy computed for vanishing momentum, which enters
  the finite temperature effective potential for the scalar field. It
  also equals the frequency of scalar field oscillations with zero $
  \vec p $. For fermions, however, the asymptotic mass is larger than
  the oscillation frequency for vanishing $ \vec p $ by a factor $
  \sqrt{ 2 } $ \cite{weldon-fermion}.}
\begin{align}  \label{mphi}
   m_\varphi ^2 
   &=  \frac{1}{16} \left (  
     3g ^2 + 4 y _ \ell ^ 2 
     g' {}^2 + 4 h_t ^2 +
     8 \lambda   
   \right ) T ^ 2          \, ,       
   \\
   \quad m_\ell ^2 &=  \frac{1}{16} \left (  3g ^2  + 4 y _ \ell ^ 2 g' {} ^2 \right ) T ^ 2
   \label{mell}
\end{align}  
where $ g $ and $ g' $ are the SU(2) and U(1) gauge couplings, $ y _
\ell = - 1/2 $ is the weak lepton hypercharge, $ h _ t $ is the top
Yukawa coupling and $ \lambda $ is the Higgs self-coupling.  Note that
the gauge field contributions to the asymptotic masses for Higgs and
leptons are equal (cf.\ Refs.~\cite{smilga,CaronHuot}). However, the
Higgs also receives important contributions from the Yukawa
interaction with the top quark and from the Higgs self-interaction, so
that $ m _ \varphi > m _ \ell $.  All other contributions can be
neglected due to the smallness of the corresponding coupling
constants.  Also, the thermal mass of the Majorana neutrinos can be
neglected.

The relevance of thermal masses for these processes was first realized
in Ref.~\cite{giudice}. There, however, the thermal mass for soft
fermionic excitations was used, by which the rate is overestimated.
In \cite{kiessig} it was argued that the lepton and/or the Higgs
momenta are soft, and that it is therefore necessary to do a Hard
Thermal Loop resummation for the Higgs and charged lepton lines. This
is correct at the edge of the threshold where the decay becomes
kinematically allowed. Since the Hard Thermal Loop gives the correct
asymptotic mass even when the external momentum is hard, the result 
of Ref.~\cite{kiessig} contains
the correct decay contribution to the rate also away from the
thresholds.
However, the
dominant contribution with this collinear kinematics is obtained
by adding 
interactions with other hard particles in the plasma, 
mediated by the exchange of  soft electroweak gauge bosons as shown 
in Fig.~\ref{fg:12}($ b $) 
\cite{anisimov}. 
In a complete leading order
calculation an arbitrary number of such interactions has to be taken into
account.  We summarize the results that were already obtained in
\cite{anisimov}, to which we refer the reader for the derivation.

\subsection{Computing the rate}

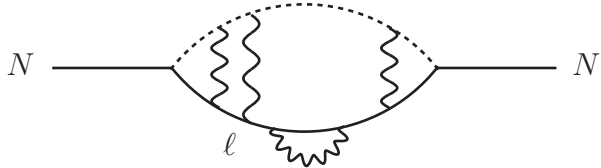
\begin{figure}[!t]
\centering
  \begin{picture}(-17,120)(220,-90)
    \SetWidth{1.0}
  \Line(105,-50)(150,-50)
  \Line(250,-50)(295,-50)
  \DashCArc(200,-90)(64,38,142){2}
  \CArc(200,-10)(64,219,-38)
  \Text(170,-83)[lb]{\normalsize{$\ell$}}
  \Photon(168,-35)(168,-65){3}{3}
  \Photon(180,-30)(180,-71){3}{3}
  \Photon(232,-35)(232,-65){3}{3}
  \PhotonArc(202,-72)(13,183,360){2}{7}
  \Text(88,-53)[lb]{\normalsize{$N$}}
  \Text(302,-53)[lb]{\normalsize{$N$}}
\end{picture}
\caption{Example for a self-energy diagram that needs to be taken into account
  in a consistent leading order calculation of the production rate via
  Eq.~(\ref{rate}).}
\label{fg:ladder}
\end{figure}

As shown in \cite{anisimov}, the kinematics described above necessitates
the inclusion of multiple soft scattering  already at leading
order. Examples for processes that must be taken into account are shown in
Fig.~\ref{fg:12}($ b $). In order to find the production rate due to this infinite
set of processes, it is most convenient to use \eqref{rate} with the
self-energy given by diagrams of the form shown in Fig.~\ref{fg:ladder} with
an arbitrary number of soft gauge boson ladder rungs or self-energy
insertions. The self-energy to be inserted in \eqref{rate} is obtained by
resumming all diagrams which respect the topology as given in
Fig.~\ref{fg:ladder}: No crossed ladder rungs or nested loops due to
self-energy insertions on the hard lines in the loop may occur, but the number
of gauge bosons and thus the total number of loops is unlimited. This means
that there is a clear mismatch between coupling constant expansion and loop
expansion: Adding more loops does not suppress the resulting rate as long as
we respect the kinematics outlined in Sec.~\ref{sc:kinematics}.

Because of this complexity, it is not possible to give an explicit result for
the retarded self-energy to be inserted in \eqref{rate}. Instead, the
diagrammatic resummation leads to an integral equation that has to be solved
numerically to obtain the rate. The retarded self-energy can be expressed by 
a 2-dimensional vector function $ \vec f $ and a scalar function $ \psi  $. 
We refer the reader to \cite{anisimov} for
details and only present the final result:
\begin{align}
  \label{gammacoll} 
      (2\pi)^3 2 k _ \parallel \frac{d\Gamma ^{ 1 \leftrightarrow  2 }
}{d ^3 k} 
      = - d ( r ) h^2
       & 
      \int \!\! \frac{d^3 p}{(2\pi)^3} 
      \frac{      f_{\rm F} (p_\parallel) 
         f_{\rm B} (k _ \parallel - p _ \parallel )}{k _ \parallel - p_\parallel} 
        \operatorname{Re} 
      \left[ 
        \frac{ k _ \parallel} { 2 p _ \| ^ 2 } \vec p _\perp \cdot \vec f
        + \frac{M _ N ^ 2 }{k _ \parallel} \psi   
      \right] \hfill 
\end{align}
where $d(r) = 2$ is the dimension of the gauge group representation
and $\vec f$ and $ \psi  $  obey the integral equations
\begin{align}
  \label{inteq_forf} 
  i\epsilon(k, \vec p) \vec f ( \vec p _\perp ) 
  - 
  \int \frac{d^2 q_\perp}{(2\pi)^2} 
  \mathcal{C} (\vec q_\perp)  
  \left[ \vec f (\vec p_\perp) - \vec f (\vec p_\perp - \vec q_\perp) \right]
  & = 2 \vec p _\perp 
  \hfill 
  ,
  \\
  \label{inteq_forpsi} i\epsilon(k , \vec p)\psi(\vec p_\perp) 
  - 
  \int \frac{d^2 q_\perp}{(2\pi)^2} 
  \mathcal{C} (\vec q_\perp)  
  \left[ \psi(\vec p_\perp) - \psi(\vec p_\perp - \vec q_\perp) \right] 
   & = 1
   ,
\end{align}
with the kernel
\begin{align} 
  {\cal C } ( \vec q _\perp ) \equiv 
  T \left [ C _ 2 ( r ) g ^ 2 
    \left ( 
      \frac{ 1 } { \vec q _\perp ^ 2 }   
      - \frac{ 1 } { \vec q _\perp ^ 2 + m^2_{\rm D} }
    \right ) 
    + y _ \ell ^ 2 g ' {} ^ 2 
    \left ( 
      \frac{ 1 } { \vec q _\perp ^ 2 }   
      - \frac{ 1 } { \vec q _\perp ^ 2 + m _{\rm D}' {}^2 }
    \right ) 
  \right ]    . 
\end{align}
Here $ m_{\rm D} $ and $ m_{\rm D} ' $ are the  
Debye masses of the SU(2) and
U(1) gauge bosons. In the Standard Model they are given by \cite{carrington}
\begin{align} 
  m_{\rm D} ^ 2= \frac{11}{6} g^2 T^2, \quad m_{\rm D} ' {} ^ 2
  = \frac{11}{6} g'^2 T^2 .
\end{align}
With $C_2(r)$ we denote the quadratic Casimir operator which for the
fundamental representation of SU(2) equals 3/4.
Finally, the quantity $\epsilon(k, \vec p)$, which gives
the difference of the energy poles of Higgs and lepton propagator, 
reads
\begin{align}\label{epsilon}
  \epsilon  (k,\vec p) 
  \equiv \frac{ M _ N ^ 2 } { 2 k _ \parallel }
  + \frac{\vec p_\perp ^2 + m_\varphi  
  ^2}{2(p_\parallel - k_\parallel)} - \frac{\vec p_\perp ^2 + m
  _ \ell 
  ^2}{2p_\parallel} .
\end{align}

The equations \eqref{inteq_forf} and \eqref{inteq_forpsi} closely resemble
those that are obtained for transverse and longitudinal photon production from
a quark-gluon plasma \cite{arnold-photon,besak,aurenche-dileptons} and can be
solved numerically by using the same idea, namely to transform them via
Fourier transformation into second-order ordinary differential equations with
boundary values. Details and some numerical results can be found in
\cite{anisimov}.

\section{
$2\rightarrow 2$ scattering}
\label{sc:2by2}

At leading order all external momenta in the $ 2 \to 2 $ scattering
processes are hard ($ k \sim T $). Furthermore, the typical angle
between the initial state momenta or between the final state momenta is
of order 1.  The masses of all external particles, including the mass
of the right-handed neutrinos, are of order $ g T $ and are neglected.
The squared tree level matrix elements $ | {\cal M } | ^ 2 $ only
depend on the Mandelstam variables $ s $, $ t $, and $ u $.  When $ |
{\cal M } | ^ 2 $ is constant or proportional to $ u/s $, the typical
momentum transfer at leading order is hard.  However when $ | {\cal M
} | ^ 2 $ is proportional to $ s/t $ or $ u/t $, there are leading
order contributions from both hard and soft momentum transfer. A naive
calculation using the tree level matrix elements would give an
infrared divergent result. Therefore one has to introduce a scale that
separates hard and soft momenta.  When the momentum transfer is hard,
one can use tree level matrix elements. For soft momentum transfer, on
the other hand, one has to use the Hard Thermal Loop resummed
propagator \cite{htl}.  Both the hard and the soft contributions
depend logarithmically on the separation scale, and only their sum is
independent.  \footnote{If the Majorana mass is not neglected, there
  are infrared divergences stemming from $ t $-channel exchange of
  soft Higgs bosons \cite{luty}.  In our calculation these divergences
  are absent since we neglect the Majorana mass, consistent with our
  power counting $ M _ N \sim g T \ll T $.}

\subsection{Hard momentum transfer}

When all external momenta and the virtualities of internal momenta are
hard, the LO production rate can simply be obtained from the Boltzmann
equation \cite{Weldon}.  More specifically, one can write the rate as
\begin{align} 
    (2\pi)^3 2 k ^ 0 & 
  \frac{d\Gamma ^{ 2\to 2  } }{d^3 k} 
   = 
   & \nonumber \\
  \sum _ { \rm processes }&  
  \int 
   \left ( \prod_{a = 1} ^3
    \frac{d^3 p_a}{(2\pi)^3 2E_a} \right )  (2\pi)^4 \delta ^ 4 (p_1
  + p_2 - p_3 - k) 
  f_1 f_2 (1 \pm  f_3) \mbox{ $ \sum $ }  {|\mathcal{M}|^2 } 
   ,
  \label{boltzmann}
\end{align}
where $f_a \equiv
f(E _ a)$ are either Bose-Einstein or Fermi-Dirac distribution
functions, and the upper/lower sign applies when particle 3 is a
boson/fermion.  Since the production rate is
defined at negligible density of Majorana neutrinos, there are no
disappearance processes and no Pauli blocking factor in
Eq.~(\ref{boltzmann}).
All momenta are hard and 
 both the Majorana mass and the  thermal masses are soft. Therefore all masses 
can be neglected 
and no propagator resummation is necessary. For the same 
reason we can put  $ E_a = | \vec p _a | $.
The invariant matrix elements squared are listed in table
\ref{tab:amplitudes}. One also has to include the $ CP $ conjugate processes,
which is done by multiplying the rate by~2. 

\begin{table}
\centering
\begin{tabular}{|l|l|l|}
\hline  
   Diagrams & Processes  &
   $\sum{|\mathcal{M}|^2 } \vphantom{\dfrac{a}{b}}$  
   \\
\hline 
      $ ( a ) $, $ ( b ) $, $ ( c ) $ 
      & $ Q_3  \bar t \to \ell  N  \vphantom{\dfrac{a}{b}}$ , 
      $\bar t \bar \ell \to \overline{ Q } _ 3 N $, $ Q_3\overline{\ell} \to  t N  
      \vphantom{\dfrac{\frac{a}{b}}{b}}$
      & $6h^2 h_t ^2 $ \\ 
\hline
  $  (d) $ & $ \ell \varphi  \to V N $   $\vphantom{\dfrac{\frac{a}{b}}{b}} $
  & $ h^2 (3g^2 + g'^2)  s/(-t)$ \\ 
\hline 
  $  (e) $ &  $ V \ell \to \overline{ \varphi}  N $ 
   $\vphantom{\dfrac{\frac{a}{b}}{b}}$ 
   & $h^2 (3g^2 + g'^2)  (-u)/s$ \\ 
\hline 
   $ (f) $ & $V \varphi  \to \bar \ell N $ $  \vphantom{\dfrac{\frac{a}{b}}{b}}$ 
   & $ h^2  (3g^2 + g'^2)   u/t  $ \\ 
\hline 
\end{tabular}
\caption{Invariant matrix elements squared and summed over spins, colors, and
  weak isospin of the initial and final states, for the $2\to 2$ 
scattering processes shown in Fig.~\ref{fg:2by2}. The results for the charge
conjugate processes are the same, and they are taken into account by
multiplying the complete $ 2\to 2 $ rate by a factor of 2. }
\label{tab:amplitudes}
\end{table} 

In order to make the phase space integration as simple as possible, it
is important to choose appropriate integration variables. We choose
them such that the denominator in $ \sum{|\mathcal{M}|^2 } $ has no
angular dependence, making the angular integrations straightforward.
All but two  integrations are performed
analytically while the two remaining integrals are done numerically.
When $ \sum{|\mathcal{M}|^2 } $ is constant or proportional to $ u/s $
we define $ q = p _ 1 + p _ 2 $. Then $s = q ^ 2 $
does not depend on any angle and many of the integrals in
\eqref{boltzmann} can be trivially performed.  On the other hand, 
when $ \sum{|\mathcal{M}|^2 } $ is 
proportional to $ s/t $, we define $ q = p _ 1 - p _ 3 $. 
In both cases we use the variables
\begin{align} 
  q _ + \equiv \frac12 \left ( q ^ 0 + | \vec q | \right  ) 
  , \qquad 
  q _ - \equiv \frac12 \left ( q ^ 0 - | \vec q | \right ) 
  ,
  \label{qpm}
\end{align} 
together with the energy of one of the colliding particles for our
non-trivial integrations.
Further details can be found in Appendix \ref{ap:details}.

For processes ($ d $) and ($ f $) which involve a fermion exchange in the $ t
$-channel the full HTL resummed propagator needs to be used as soon as the
momentum transfer becomes soft, while for a hard momentum a bare propagator is
sufficient. A successful implementation was first given by Braaten and Yuan
\cite{Yuan} and relies on introducing a cut for the square of the spatial
momentum transfer.  Here we proceed slightly differently from
\cite{Yuan,moore-testing,arnold-complete,gravitino,axino}. 
We introduce a cut for the square
of the {\em transverse} momentum $ \vec q _\perp $. By Eq.~(\ref{qperp}) $
\vec q _\perp ^ 2 $ is determined by $ q _ + $ and $ q _ - $. We take
\begin{align}
  g T \ll q _ { \rm cut } \ll T, 
  \label{qcut}
\end{align}  
and integrate the tree level $ 2 \to 2 $ scattering matrix elements
only over transverse momenta with $ \vec q _\perp ^ 2 > q  ^ 2 _ {\rm cut }
$.  Due to (\ref{qcut}) and (\ref{qperp}) this condition turns out to be
equivalent to $ - q ^ 2 > q _ {\rm cut } ^ 2 $.  The region 
 $  \vec q _\perp ^ 2
< q  ^ 2 _ \text {cut } $ is discussed in Sec.~\ref{sc:soft}. There we will 
see why it is convenient to have a cut in transverse
momenta only: Then we can use the sum rule for the HTL resummed
propagator obtained in Appendix \ref{ap:sumrule}.

We first perform all integrations except over $ q _ 
+ $ and $ q _ - $,
see Eq.~(\ref{t-channel}). The resulting integrand diverges
like $ 1/\vec q ^ 2 $ for $ q  \to 0$,
\begin{align} 
  \left [  \frac{d\Gamma}{d^3 k} \right ]  _ {\rm hard} 
  = \int  _ 0 ^ { k ^ 0 } d q _ 
  + \int  _ { -{\infty  } } ^ 0 d q _ - \frac{ F (
    q _ + , q _ - )  } { \vec q ^ 2 } 
  \,
  \Theta  ( \vec q _\perp ^ 2 - q _ {\rm cut } ^ 2 ) 
  \label{hard}
  ,
\end{align}
which can be seen from Eqs.~(\ref{t-channel}) and (\ref{sdurcht}).   
Now we subtract and add the $ q _ + $, $ q _ - \to 0 $ limit of $ F $.
This way we decompose (\ref{hard}) into a piece which is
finite when $ q _ {\rm cut } \to 0 $, and one which is singular in this limit,
\begin{align} 
  \left [  \frac{d\Gamma}{d^3 k} \right ] _ {\rm hard } 
  =
   \left [ \frac{d\Gamma}{d^3 k}   \right] _ {\rm hard, finite }  
   +
   \left [  \frac{d\Gamma}{d^3 k}  \right ] _ {\rm  hard, singular} 
   .
\end{align} 
In the first term on the RHS we can put $ q _ {\rm cut } = 0 $
which gives
\begin{align} 
   \left [  \frac{d\Gamma}{d^3 k} \right ] _ {\rm hard, finite } 
   = 
   \int _ 0 ^ { k ^ 0 } d q _ + \int _ { -{\infty  } } ^ 0 d q _ - \frac{ F (
    q _ + , q _ - )  - F ( 0, 0 )  } { \vec q ^ 2 } 
  \label{hardfinite}
  . 
\end{align} 
$ F $ is obtained by combining Eqs.~(\ref{boltzmann}),
(\ref{t-channel}) together with
Eqs.~(\ref{E1integrald}) and (\ref{E1integralf}).
  The remaining integrals have to be done
numerically.  The subtracted
term is singular for $ q _ {\rm cut } \to 0 $, but it can be computed
analytically. We find equal contributions for the processes $ ( d ) $
and $ ( f ) $. We also have to include the charge conjugated
processes, which gives an additional factor 2. The complete singular
hard contribution is
\begin{align} 
   2 k ^ 0 ( 2 \pi  ) ^ 3 
   \left [  \frac{d\Gamma}{d^3 k}  \right ] _ {\rm hard, singular}
   =    
   \frac{h ^ 2}{2 \pi  } m _ \ell ^ 2  f _ { \rm B } ( k ^ 0 ) 
   \ln \left ( \frac{ 2 k ^ 0  } { q _ {\rm cut }  }
   \right ) 
   \label{hardsingular}
\end{align} 
where $ m _ \ell $ is the asymptotic mass of the leptons  (\ref{mell}).

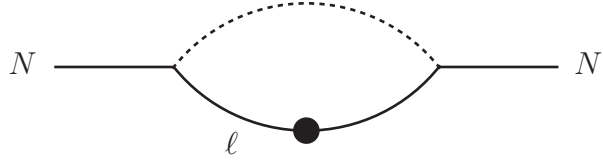
\begin{figure}[!t]
\centering
  \begin{picture}(-17,120)(220,-90)
    \SetWidth{1.0}
  \Line(105,-50)(150,-50)
  \Line(250,-50)(295,-50)

  \DashCArc(200,-90)(64,38,142){2}

  \CArc(200,-10)(64,219,-38)
  \Text(170,-83)[lb]{\normalsize{$\ell$}}

  \Text(88,-53)[lb]{\normalsize{$N$}}
  \Text(302,-53)[lb]{\normalsize{$N$}}
 \Vertex(200,-74){5}
\end{picture}
\caption{One-loop self-energy diagram of the Majorana neutrino. 
The black dot indicates that the lepton propagator is HTL resummed.}
\label{fg:HTL}
\end{figure}

\subsection{Soft momentum transfer}
\label{sc:soft}

Next we compute the contribution with soft transverse momentum
transfer, $ \vec q _\perp ^ 2 < q _ { \rm cut } $. 
Here we make use of \eqref{rateRH} to compute the rate.  
The leading order
contribution is given by the diagram in Fig.~\ref{fg:HTL}, where the
internal lepton line is soft. The leading order self-energy for soft
fermions is given by a 1-loop diagram with a gauge field propagator
and a hard loop momentum, a so-called Hard Thermal Loop (HTL).  It is
of the same order as the lepton momentum. Therefore one has to use the
HTL resummed propagator as indicated by the thick
blob in Fig.~\ref{fg:HTL}.  
The soft contribution to the rate depends on the separation scale $ q
_ {\rm cut } $, and adding the soft contribution and the hard singular
gauge boson scattering rate (\ref{hardsingular})
the separation scale drops out.
 
There is a similar contribution to thermal photon
production which has been computed in Refs.~\cite{kapusta,baier}. There
it turned out that the result, which was found numerically, could be
written in a simple analytic form \cite{arnold-complete,kapusta}, 
despite the fact that 
the HTL self-energy is a rather non-trivial function. Here we
trace the reason why the complicated integrals in
\cite{arnold-complete,kapusta,baier} have such a simple result back to
a sum rule which is satisfied by the HTL resummed fermion propagator.

We must
compute the imaginary part of the self-energy from Fig.~\ref{fg:HTL}
with a HTL resummed lepton propagator $ S $ which is given by
Eq.~(\ref{shtl}).  The Higgs momentum is hard, so one can use the tree
level Higgs propagator.  In the imaginary time formalism the self-energy reads
\begin{align}
  P _ {\rm L} \Sigma  (k) = d ( r ) h^2  P _ {\rm L} T
  \sum_{q^0} 
  \int\frac{d^3 q}{(2\pi)^3} \Delta(k - q) 
  S
  (q) 
\end{align}
with the free Higgs
propagator $ \Delta ( p ) = -1/p^2 $. After performing the sum  over
fermionic Matsubara frequencies $ q ^ 0 $ one can analytically continue
$ k ^ 0 $ to real values.

One can decompose the fermion propagator according to  $ S = S ^ \mu  \gamma  _
\mu  $. In the calculation of the rate only   the discontinuity        
\begin{align} 
  { \rm disc } S ( q ) \equiv S ( q ^ 0 + i 0 ^ +, \vec q ) - S ( q ^ 0
  + i 0 ^ - , \vec q ) 
  \label{disc}
\end{align} 
across the real $ q ^ 0$-axis (with real $ \vec
q $) of the  $ + $ component $ S _ + \equiv S _ 0
+ S _ z = S ^ 0 - S ^ z $ enters.
The integration over $ q ^  z $ is done using the sum rule (see Appendix
\ref{ap:sumrule})
\begin{align} 
  \int \frac{ d q ^ z } { 2 \pi  }\,  
    { \rm disc } S _ + ( q ^ 0= q ^ z , \vec q _\perp )  
    =  
    \frac{ i } { 2 } \: \frac{ m _ \ell ^ 2} { \vec q _\perp ^ 2 + m _ \ell ^ 2 } 
    \label{sumrule}
     .
\end{align}
Finally one can integrate over the transverse momenta with $ \vec q _\perp ^ 2
< q _ {\rm cut } ^ 2 $ which gives
\begin{align} 
   2 k ^ 0 ( 2 \pi  ) ^ 3 
   \left [  \frac{d\Gamma}{d^3 k}  \right ] _ {\rm soft}
   =    
   \frac{h ^ 2}{2 \pi  } m _ \ell ^ 2 f _ { \rm B } ( k ^ 0 ) 
   \ln \left ( \frac { q _ {\rm cut }  } { m _ \ell } 
   \right ) 
   .
   \label{soft}
\end{align} 
Adding the singular hard contribution (\ref{hardsingular}) and
Eq.~(\ref{soft}) the separation scale $ q _ {\rm cut } $ drops out.

\section{Numerical results}
\label{sc:results}

The total rate of the  $ 2 \to 2 $ processes can be written  
as
\begin{align} 
  \Gamma^{2\rightarrow 2} 
  = 
    \frac{ h^2 T^4 }  { 1536\pi  } 
    \left\{
    h _ t  ^2  c _ {Q } +
    (3g^2 + g'^2) 
    \left [  
      \ln \left ( \frac{1}{ 3 g^2 + g'^2 } \right ) 
      + c _ V 
      \right ] 
    \right\} 
  \label{222formel}
   ,
\end{align} 
where $ h _ t $ is the top Yukawa coupling.  We determined the
constants $ c _ Q $ and $ c _ V $ numerically, with the results
\begin{align} 
  c _ Q & = 2.52 \,,
  \\
  c _ V & = 3.17 \,
  .
\end{align} 
For the $ 1 \leftrightarrow 2 $ processes there is no such simple expression,
since the rate is a complicated function of the ratios of thermal and Debye masses.
\begin{figure}[t]
  \centerline{
    \epsfxsize=11cm
    \epsffile{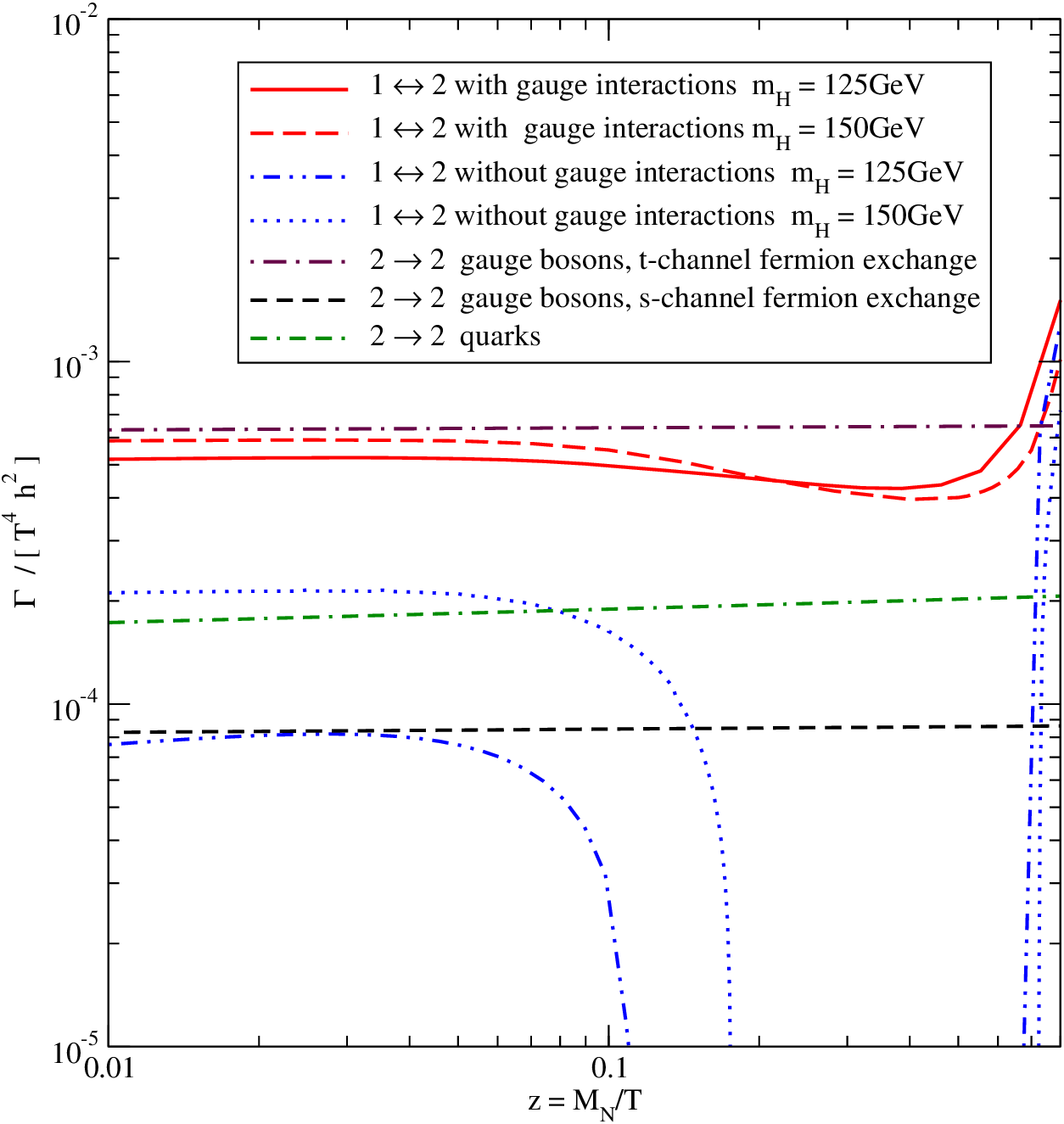}
  } 
  \caption  {Dependence of the production rate on $ z $  for $ M _ N = 10 ^ 7 $
GeV.  
    \label{fg:gammavsz}}
\end{figure}

To obtain numerical results for the rate we have to specify several
parameters.  The mass $ M _ N $ of the Majorana neutrino and its Yukawa
couplings are unconstrained by low-energy neutrino physics. If not stated
otherwise we have chosen the exemplary value $M_N = 10^7$ GeV and we always
plot our rates divided by $ h^2$.  The SM couplings are evaluated at the scale
$\mu = 2\pi T$ using the 1-loop renormalization group equations~\cite{RG}. The
Higgs self-coupling $ \lambda $ is determined by the zero temperature Higgs
mass $ m _ H $ for which we have considered the two values $ 125 $ and $ 150
$~GeV.

In Fig.~\ref{fg:gammavsz} we show rates for the various processes as a
function $ z \equiv M _ N/T $. The 
$ 1 \leftrightarrow 2 $ processes show the characteristic behavior
found in \cite{anisimov}. Without gauge interaction (dotted and dot-dot-dashed
lines)
there is the Higgs decay at very high temperature and the inverse
decay at lower temperature, and there is a gap in between where no
process is kinematically allowed. Adding soft gauge interactions
closes this gap.  Even when the Higgs decay is allowed, gauge interactions
enhance the rate by almost a factor 3.  When the inverse $ N $ decay
is kinematically allowed, which happens close to $ z = 1 $, the effect
of gauge interactions becomes small, and our result for the complete $
1 \leftrightarrow 2 $ rate is approximately equal to the inverse decay
rate of the right-handed neutrinos.  This indicates
that our $ 1 \leftrightarrow  2 $ result
 which assumed the Majorana neutrino mass to be
soft ($ M _ N \sim g T $) would smoothly match onto the result for
hard ($ M _ N \sim T $) Majorana mass, for which the soft gauge
interactions are suppressed. However, we cannot expect our $ 2 \to 2 $ results
to be valid in this region since they were obtained by neglecting $ M _ N $.

\begin{figure}[t]
  \centerline{
    \epsfxsize=9cm
    \epsffile{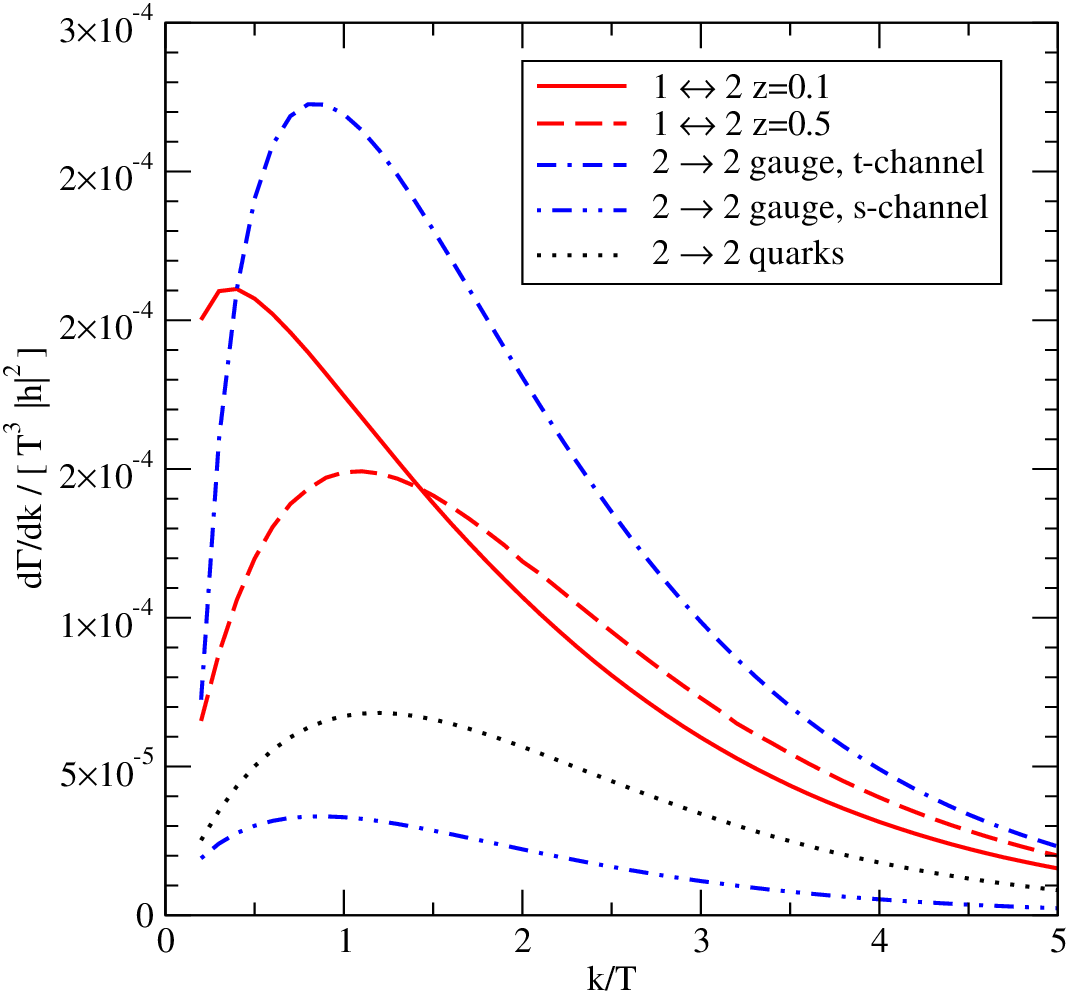}
  }
  \caption  {Momentum spectrum of the produced right-handed neutrinos for $ M
    _ N = 10
    ^ 7 $ GeV and $ m _ H = 125 $ GeV.
    $ s $- and $ t $-channel refers to the exchanged fermion.
    \label{fg:spectrum}}
\end{figure}

The contribution from $ 2\to 2 $ gauge boson scattering with $ t
$-channel fermion exchange is about the same size as from the $ 1
\leftrightarrow 2 $ scattering with soft gauge interaction.
This shows that gauge bosons, both real and virtual, are essential for
the production, ignoring them would strongly underestimate the
rate.  The quark scattering contributions are smaller than
the gauge boson scattering by more than a factor 3, despite the large
top Yukawa coupling.  The rates of the $ 2 \to 2 $ processes depend
only logarithmically on $ z $ through the running of the gauge
coupling, since we have neglected the Majorana mass in the $ 2\to 2 $
matrix elements

Refs.~\cite{pilaftsis,hahn} find that the $ 1 \leftrightarrow 2 $
processes without soft gauge interaction give a much smaller rate at
high $ T $ than the $ 2\to 2 $ scattering.  This is because they do
not include thermal masses in the $ 1 \leftrightarrow 2 $ rates which
are, however, crucial at high temperature. Thermal masses were
included in Ref.~\cite{giudice}. The resulting rate is about 60\%
larger than ours.  This could be related to the fact that
\cite{giudice} uses the mass for soft fermion excitations instead of
the asymptotic lepton mass $ m _ \ell $ which is larger by a factor $
\sqrt{2} $, thus overestimating the rate for Higgs decay.
In Ref.~\cite{giudice} it was found that at very high $ T $ the
complete $ 2 \to 2 $ scattering rate is approximately equal to the
rate due to Higgs decay without soft gauge interaction (see
Eq.~(\ref{decay})).  \footnote{If one wants to compare the
  individual contributions of \cite{giudice} with ours one has to be careful: In
  the present paper $ s $- or $ t $-channel refers to the fermion
  exchange, while in Ref.~\cite{giudice} it refers to the Higgs
  exchange.} Here we find that the former gives a much larger
contribution. This discrepancy cannot only be due to the use of the
soft fermion mass in Ref.~\cite{giudice}. Another source of
discrepancy could be that in \cite{giudice,pilaftsis} the soft fermion
exchange is dealt with differently from our complete leading order treatement.
\begin{figure}[t]\vspace{-0cm}
  \centerline{
    \epsfxsize=11cm
    \epsffile{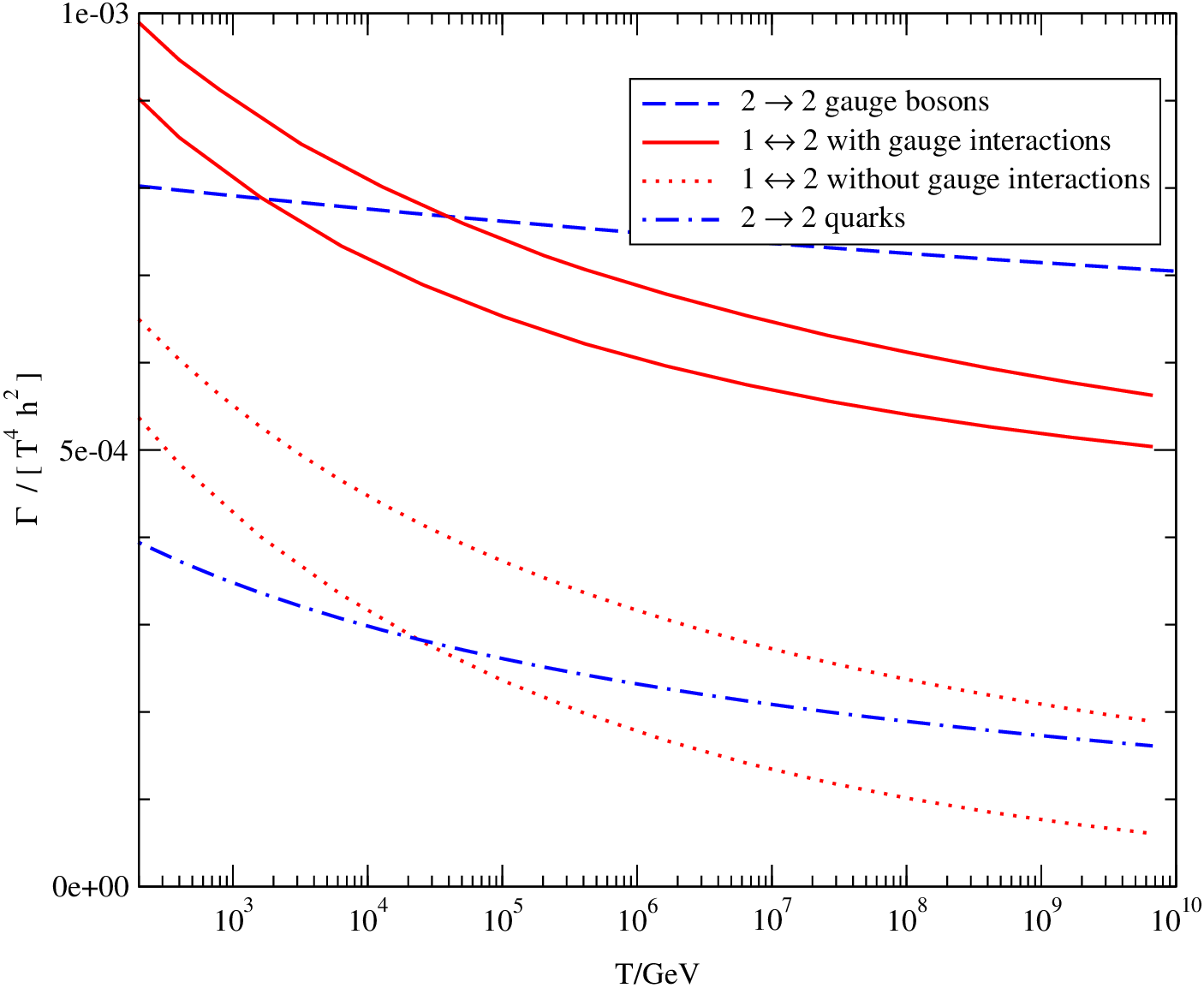}
  }
  \caption  {Production rate of massless
    right-handed neutrinos
     as a function of the temperature. The upper and lower $ 1 \leftrightarrow  2
     $ curves are for $ m _ H = 150 $ and $ 125 $~GeV, 
     respectively.
    \label{fg:temperature}}
\end{figure}
Both 
\cite{pilaftsis} and 
\cite{giudice} find that at high temperatures the $ 2 \to 2 $ gauge
boson scattering dominates over the quark scattering by about a factor
2. Here we find that the gauge boson scattering dominates by a much
larger factor of about 5. 
 
Fig.~\ref{fg:spectrum} shows the momentum spectrum of the produced
neutrinos. For $ 2 \to 2 $ processes the typical momentum is near $ T
$. This is also the case for the $ 1 \leftrightarrow  2 $ contribution at $ z =
0.5 $, where without gauge interaction no decay or inverse decay is
allowed. However in the high temperature regime where the Higgs decay is
allowed the spectrum is much more infrared. This does not invalidate our
calculation, because it is still applicable  when the right-handed neutrinos are
soft. 

In Fig.~\ref{fg:temperature} we show the temperature dependence of the
production rate for vanishing Majorana mass. This is of particular
importance for leptogenesis which occurs far from thermal equilibrium
at relatively low temperature, for instance below the electroweak
phase transition \cite{asaka-bau}. For all temperatures between $ 100
$ and $ 10 ^{ 10 } $ GeV the rates for $ 1 \leftrightarrow 2 $
scattering plus soft gauge interaction and the gauge boson scattering
are of similar size, and they dominate over the quark scattering.
At lower temperatures  the $ 1 \leftrightarrow 2 $ scattering
gives the largest contribution, while at higher $ T $ the $ 2\to
2 $ gauge boson scattering takes over. Numerical results for the $ 1
\leftrightarrow  2 $ production rate are also shown in Tab.~\ref{tab:rates}.

Another interesting result of this paper is the Higgs mass dependence
of the $ 1 \leftrightarrow 2 $ rate which is displayed in
Figs.~\ref{fg:gammavsz} and \ref{fg:temperature}.  Without gauge
interactions we observe a strong suppression in the region where the
Higgs can decay, while the rate gets enhanced in the region of inverse
$ N $-decay.  When gauge interactions are included, the relative $ m _
H $-dependence is much weaker.

\begin{table}
\centering
\begin{tabular}{c | c}
    $ T $/GeV &  $   \Gamma  /(h ^ 2 T ^ 4 )   $ \\
\hline 
$ 1.00 \times   10 ^ 2 $   &  $ 9.5 \times 10 ^ {-4} $ \\
$  1.60 \times 10 ^ 3  $    & $ 7.9 \times 10 ^ {-4}$\\ 
$  2.56 \times 10 ^ 4 $  &   $ 6.9  \times 10 ^ {-4}$\\
$  4.10 \times 10 ^ 5 $   &  $6.2  \times 10 ^ {-4}$\\
$  6.55 \times 10 ^ 6 $  &   $5.7 \times 10 ^ {-4}$ \\
$  1.05 \times 10 ^ 8 $ &    $ 5.4 \times 10 ^ {-4}$ \\
$  1.68 \times 10 ^ 9 $  &   $5.1  \times 10 ^ {-4}$\\
$  6.71 \times 10 ^ 9 $   &  $5.0 \times 10 ^ {-4} $\\
\end{tabular}
\caption{Production rate $ \Gamma  $  of massless right-handed neutrinos from 
 $ 1 \leftrightarrow  2 $  scattering including gauge
  interactions for $ m _ H = 125 $~GeV.}
\label{tab:rates}
\end{table} 

\section{Summary and Conclusions}
\label{sc:conclusions}

In this paper we have computed the complete leading order production
rate of right-handed neutrinos at high temperature, assuming that
their mass $ M _ N $ is soft, $ M _ N \sim g T $. There are two types
of contributions.  The first is due to $ 1 \leftrightarrow 2 $
scattering processes with additional soft gauge interactions. They are
characterized by the near collinearity of the participating hard
particles, and we have computed them before. Here we have studied
their dependence on the zero temperature Higgs mass, which turned out
to be weak when soft gauge interactions are included. We
have completed the leading order calculation by also including $ 2 \to
2 $ scattering processes.  The $ 2 \to 2 $ processes involving 
electroweak gauge bosons were consistently computed for the first time
by properly treating the effects of the hot plasma on  soft lepton exchange.

The gauge boson $ 2 \to 2 $ rate is of similar size as the complete $
1 \leftrightarrow 2 $ rate.  Both are significantly larger than the
top quark scattering contribution despite the large top Yukawa
coupling.  At lower temperatures the $ 1 \leftrightarrow  2 $ processes 
give the largest 
contribution, while at higher temperature the $ 2 \to 2 $ rate takes over.
Our main conclusion is thus that processes
involving gauge interactions are the most important ones for producing
right-handed neutrinos at high temperature.  At higher order the
processes discussed in this paper also contribute to the
CP asymmetry.
Thus one should expect gauge interactions to play a crucial role also 
in that context where they have not been taken
into account so far \cite{giudice,kiessig-htl1}.
 
\vspace{.5cm} 
\noindent{\bf Acknowledgments} DB thanks Bj\"orn Garbrecht and 
Mikko Laine for useful discussions and  suggestions, and Mikko Laine
for pointing out an error in an earlier version of this paper.
This work was supported in part through
the DFG funded Graduate School GRK 881.


\appendix 
\renewcommand{\theequation}{\thesection.\arabic{equation}}

\section{Details of the calculation 
of the $2\rightarrow 2$ rate}   
\label{ap:details}
\setcounter{equation}0

The phase space integrations in Eq.~(\ref{boltzmann}) are done as
follows. Following \cite{moore-testing,arnold-complete} we introduce
the 4-momentum $ q $ of an exchanged virtual particle 
such that $ q ^ 2 $ appears in the denominator of the
squared matrix elements. After doing some angular integrals
we end up with  non-trivial integrations
over $ q _ + $, $ q _ - $, which are defined in
Eq.~(\ref{qpm}), and the energy $ E' $ of one of the
colliding particles.  For each process the products of Bose and Fermi distribution functions in
Eq.~(\ref{boltzmann}) are written in the form 
\begin{align}
  f _ 1 f _ 2 \left [ 1 \pm f _ 3 \right ]
  = f _ { \rm F } ( k ^ 0 )\widetilde f   \, \widehat   f
  \label{fprod}
  ,
\end{align}
where $ k ^ 0 $ is the energy of the produced right-handed neutrino. $ \widetilde
f $ and $ \widehat f $ are both linear combinations of Bose and Fermi
distribution functions which are process dependent. They are chosen such that
only $ \widehat f $ depends on $ E ' $.  This greatly simplifies the integrals
over $ E ' $ which are then performed analytically.  
In order to arrive at Eq.~(\ref{fprod}) we use the identity
\begin{align}
 f _ { \rm B } ( E _ 1 ) f _ { \rm B } ( E _ 2 ) = f _ {\rm B } ( E _ 1 + E _ 2 ) 
 \left [ 1 + f _ { \rm B } ( E _ 1 ) + f _ { \rm B } ( E _ 2 ) \right ]
 ,
\end{align} 
and the corresponding  relations for Fermi distributions which are obtained 
using 
\begin{align} 
  f _ { \rm F } ( \omega   ) = - f _ { \rm B } ( \omega   + i \pi  T ) 
  ,
\end{align}
together with
\begin{align}
    f _ {\rm B }  ( - \omega    ) = - [ 1 + f _ {\rm B }  ( \omega   ) ] 
    .
\end{align} 
The remaining integrations
over $ q _ + $ and $ q _ - $ are done numerically. In the processes with
fermion exchange in the $ t $-channel we perform a subtraction in the
integrand to isolate the singular contribution.

\subsection{$ s $-channel parametrization}

For gauge boson scattering with $ s $-channel fermion exchange, where
$ \sum | { \cal M } | ^ 2 $ is proportional to $ u/s $, and for  quark scattering
where it is constant, we choose
\begin{align} 
  q =  p_3 +  k 
\end{align} 
so that $ s = q ^ 2 $, and $ E '  = E _ 2 $.
For $ n = 0 $ or $ n = 1 $ we obtain
\begin{align} 
  \int  \left [  \prod_{a = 1} ^3
    \frac{d^3 p_a}{(2\pi)^3 2E_a} \right ]  &
(2\pi)^4 \delta (p_1   + p_2 - p _ 3 - k ) 
 \widetilde f   \, \widehat   f \left ( \frac{ -u } { s } \right ) ^ n 
  \nonumber \\
  &
  = 
  \frac{ 1 } { 32 \pi  ^ 3 k ^ 0} 
  \int _ { k ^ 0 } ^ {\infty  } d q _ + \int _ 0 ^ { k ^ 0 }  d q _ - \widetilde f
  \int _ { q _ -   } ^{ q _ + } d E _ 2 \widehat f
  \left ( 
    \frac{ \langle -u \rangle } { s } 
  \right )  ^ n 
  \label{s-channel}
\end{align} 
with 
\begin{align} 
  \frac{ \langle -u \rangle } { s } 
  =
  \frac 1 {  \vec q ^ 2 } 
   \left (   \frac 1 2 \left ( q _ 0 ^ 2 + \vec q ^ 2 \right ) 
     - q ^ 0 ( E _ 2 + k ^ 0 ) + 2 E _  2 k ^ 0 \right )
\end{align} 
and $ q ^ 0 = q _ + + q _ - $, $  q ^ 2  = 4 q _ + q _ - $, and $ | \vec
q |  =  q _ + - q _ - $. The angular brackets indicate that 
the Mandelstam variable $ t $ has been averaged 
over the azimuth of $ \vec p _ 2 $ 
with respect to $ \vec q $.
We have to distinguish two cases:
\subsubsection{quark scattering}
All 
external particles are fermions, and we use 
\begin{align}
    \widetilde f &= f _ { \rm B } ( E _ 3 + k ^ 0 )  + f _ { \rm F } ( E _ 3 )  
    \label{ftildeabc}
    \, ,
    \\
    \widehat   f &= 1 - f _ { \rm F } ( E _ 1 ) - f _ { \rm F } ( E _ 2 ) 
    \label{fhatabc}
    \,,
\end{align}
where $ E _ 3 = q ^ 0 - k ^ 0 $ and $ E _ 1 = q ^ 0 - E _ 2 $.  $  \sum | {
  \cal M } | ^ 2 $ 
is constant  which corresponds to $ n = 0 $ in Eq.~(\ref{s-channel}) and we find 
\begin{align}
   \int _ { q _ -   } ^{ q _ + } d E _ 2 \widehat f
   =  | \vec q |   +2 T \left [ 
     \log \left(1 + e^{-q _ + /T}\right) - \log \left(1 + e^{-q _ -/T}\right)
   \right ] 
   \,.
\end{align}

\subsubsection{$ V \ell \to \overline{ \varphi}  N $  }

Here we have both external fermions and bosons, so that 
\begin{align}
    \widetilde f &= f _ { \rm F } ( E _ 3 + k ^ 0 )  + f _ { \rm B } ( E _ 3 )  
    \label{ftildee}
    \,,
    \\
    \widehat   f &= 1 + f _ { \rm B } ( E _ 1 ) - f _ { \rm F } ( E _ 2 ) 
    \label{fhate}
    \,.
\end{align}
The  $ E _ 2$-integral in Eq.~(\ref{s-channel}) gives
\begin{align} 
    \int _ { q _ -   } ^{ q _ + }  d E _ 2 \widehat f \,
    \frac{ \langle -u \rangle } { s }  
    = &\frac{ | \vec q | } { 2 } 
    -
    \frac{ T } { | \vec q | } 
   \Big[  ( k ^ 0 - q _ + ) 
         \Big( \ln \left ( 1 - e ^{ - q _ + / T } \right ) 
             - \ln \left ( 1 + e ^{ - q _ - /T } \right )
             \Big ) 
    \\ \nonumber 
   &  \qquad \qquad   +  ( k ^ 0 - q _ - ) 
   \Big ( \ln \left ( 1 - e ^{ - q _ - /T } \right ) 
       - \ln \left ( 1 + e ^{ - q _ + / T } \right ) 
       \Big ) 
\Big ]  
\\ {}-
  \frac{ T ^ 2 } { \vec q ^  2}  ( q ^ 0 - 2 k ^ 0 ) 
  &
    \Big [ {\rm Li}_2  \left ( e ^{ - q _ + / T } \right ) 
      - {\rm Li}_2  \left ( e ^{ - q _ - /T } \right )
      - {\rm Li}_2  \left ( - e ^{ - q _ + /T } \right ) 
      + {\rm Li}_2  \left ( - e ^{ - q _ - / T } \right )
      \Big ]  \nonumber 
      ,
\end{align}
where $ {\rm Li}_2  $ is the dilogarithm.

\subsection{$ t $ -channel parametrization: hard contribution}

For a process with a fermion exchange in the $ t $-channel we choose 
\begin{align} 
    q =  p_1 -  p_3
    \,,
\end{align} 
so that $ t = q ^ 2 $, and $ E '  = E _ 1 $.
For the phase space integral in
Eq.~(\ref{boltzmann}) we obtain
\begin{align} 
  \int  \left [  \prod_{a = 1} ^3
    \frac{d^3 p_a}{(2\pi)^3 2E_a} \right ]  
  &
  (2\pi)^4 \delta (p_1   + p_2 - p _ 3 - k) 
  \widetilde f   \, \widehat   f \left ( \frac{ s } { -t } \right ) ^ n 
 \nonumber \\
  &
  = 
  \frac{ 1 } { 32 \pi  ^ 3 k ^ 0 } 
  \int _ 0 ^ {k ^ 0  } d q _ + \int _ { - {\infty  } }  ^ 0 d q _ - \widetilde f
  \int _ { q _ +   } ^{ {\infty  }  } d E _ 1 \widehat f
  \left (  \frac{ \langle s \rangle } { -t } 
  \right )  ^ n 
    \label{t-channel}
\end{align} 
where we need the cases $ n = 0 $ and $ n = 1 $. Furthermore,
\begin{align}
  \frac{ \langle s \rangle } { -t } = 
  \left (  \vec q ^ 2 + 2 E _ 1 k ^ 0 -  q ^ 0 ( E _ 1 + k ^ 0 ) 
      + \frac{ q ^ 2 } { 2 } \right ) /\vec q ^ 2 
    \label{sdurcht}
\end{align} 
with $ q ^ 0 $, $ q ^ 2 $, and $ \vec q ^ 2  $ as in
Eq.~(\ref{s-channel}). The transverse momentum transfer is given by
\begin{align}
  \vec q _\perp ^ 2 = - \frac{ q ^ 2 } { k _ 0 ^ 2 } ( k _ 0 - q _ + ) 
  ( k _ 0 - q _ - )
  \label{qperp}
  \,.
\end{align} 
Thus for small $ q $ we have $ \vec q _\perp ^ 2 \simeq - q ^ 2 $.  Now the
angular brackets indicate an average over the azimuth of $ \vec p _ 1 $
with respect to $ \vec q $.  During $ E _ 1 $-integration $ E _
3 = E _ 1 - q ^ 0 $ changes, while $ E _ 2 = k ^ 0 - q ^ 0 $ remains fixed.  Again
we have to distinguish two cases:

\subsubsection{$ \ell \varphi  \to V N $ }

In this case
\begin{align}
   \widetilde f =& 1 + f _ { \rm B } ( E _ 2 ) - f _ { \rm F } ( q ^ 0 ) 
   \label{ftilded}
   ,
   \\
   \widehat f = & f _ { \rm F } ( E _ 1 ) + f _ { \rm B } ( E _ 3 ) 
   .
\end{align} 
The matrix element squared is proportional to $ s/t $, so that we have the 
$ E _ 1 $-integral
\begin{align} 
  \int _ { q _ + } ^ {\infty  } d E _ 1 \hat{f} 
  \, \frac{ \langle s \rangle } { -t }  \, = \, &
  \frac{ T } {  | \vec q | } ( k ^ 0 - q _ - ) 
  \Big [ \ln \left ( 1 + e ^{ - q _ + /T } \right ) 
  -      \ln \left ( 1 - e ^{ q _ - /T } \right ) 
  \Big ] 
  \nonumber \\
  & + \frac{ T ^ 2 } { \vec q ^ 2 } ( 2 k ^ 0 - q ^ 0 ) 
  \Big [ - {\rm Li}_2  \left ( - e ^{ -q _ + / T } \right ) 
        + {\rm Li}_2  \left ( e ^{ q _ - / T } \right ) 
  \Big ] 
  \label{E1integrald}
  \,.
\end{align}
For the subtraction in Eq.~(\ref{hardfinite}) we need the small $ q $ limit of
(\ref{E1integrald}) for which we find
\begin{align} 
  \int _ { q _ + } ^ {\infty  } d E _ 1 \hat{f} 
  \, \frac{ \langle s \rangle } { -t }  \, = \,
  \frac{ \pi  ^ 2 T ^ 2 k ^ 0 } { 2 \vec q ^ 2 } + O ( q ^{ -1 } )
  \label{E1integraldsoft}
  \,.
\end{align}

\subsubsection{$V \varphi  \to \bar \ell N $ }

Now we have
\begin{align}
   \widetilde f =  & 1 + f _ { \rm B } ( E _ 2 ) - f _ { \rm F } ( q ^ 0 ) 
   \label{ftildef}
   ,
   \\     
   \widehat f = & f _ { \rm B } ( E _ 1 ) + f _ { \rm F } ( E _ 3 ) 
   .
\end{align} 
The matrix element squared is proportional to $ u/t = -1 - s/t $, 
and the integration over $ E _ 1 $ yields
\begin{align}
  \int _ { q _ + } ^ {\infty  } 
  d E _ 1 \hat  f \frac{ \langle u \rangle } { t } \, =\, &
  \frac{ T } { | \vec q | } ( k ^ 0 - q _ + ) 
  \Big [ - \ln \left ( 1 - e ^{ - q _ + / T } \right ) 
  + \ln \left ( 1 + e ^{ q _  - / T } \right ) 
  \Big ] 
  \nonumber\\
  & {} + \frac{ T ^ 2 } { \vec q ^ 2 } ( 2 k ^ 0 - q ^ 0 ) 
  \Big [ {\rm Li}_2  \left ( e ^{ - q _ + /T } \right ) 
  - {\rm Li}_2  \left ( - e ^ { q _ - / T } \right ) 
  \Big ] 
  \label{E1integralf}
  .
\end{align}
The term which is subtracted in Eq.~(\ref{hardfinite}) is the same as 
for the processes $ \ell \varphi  \to V N $ because 
\begin{align} 
  \int _ { q _ + } ^ {\infty  } d E _ 1 \hat{f} 
  \, \frac{ \langle u \rangle } { t }  \, = \,
  \frac{ \pi  ^ 2 T ^ 2 k ^ 0 } { 2 \vec q ^ 2 } + O ( q ^{ -1 } )
  \label{E1integralfsoft}
\end{align} 
is the same as (\ref{E1integraldsoft}).

\section{Sum rule for HTL resummed fermion propagotor}
\label{ap:sumrule}
\setcounter{equation}0

Here we show that the Hard Thermal Loop (HTL) resummed 
propagator for  massless fermions
\begin{align}
  S ( q ) = \frac{ -1 } { \cancel{q} - \Sigma  _ { \rm HTL } ( q ) } 
  \label{shtl}
\end{align} 
satisfies the sum rule (\ref{sumrule}). 
$ \Sigma _ { \rm HTL } = \Sigma _ { \rm HTL } ^ \mu \gamma _ \mu $ is
the HTL self-energy \cite{weldon-fermion,klimov}. Its components can be
written as
\begin{align} 
   \Sigma _ { \rm HTL } ^ 0 ( q )& = \frac{ m _ \ell ^ 2 } { 4 } q ^ 0
   \int _ {-1 } ^ 1 \frac{ d x } { ( q ^  0 ) ^ 2 - x ^ 2 \vec q ^ 2 } 
   \label{s0}
   \\
   \vec \Sigma   _ { \rm HTL } ( q ) &= 
   \frac{ \vec q } { \vec q ^ 2 } 
   \left ( q ^ 0 \Sigma  ^ 0 _ { \rm HTL }
     - \frac{ m _ \ell ^ 2 } { 2 } \right ) 
   \label{si}
\end{align}
with the asymptotic thermal mass $ m _ \ell $ (see Eq.~(\ref{mell})).  The
sum rule (\ref{sumrule}) is similar to the one for the gauge field propagator
found in Ref.~\cite{aurenche-sumrule}. The reason why the complicated HTL
resummed propagator gives such a simple result has been identified in
\cite{caron-huot}. Due to causality, the propagator is analytic not only in the
upper and lower half of the complex $ q ^ 0 $-plane, but also when an
imaginary light-like vector is added to $ q $.  
This also holds for the fermion propagator as can be easily seen from 
Eqs.~(\ref{s0}) and (\ref{si}). Therefore one can move the
integration contour in Eq.~(\ref{sumrule}) for the first term in
Eq.~(\ref{disc}) into the upper $ q ^ z $-half plane and for the second one
into the lower half plane. The integrand $ S _ + $ falls off like $ 1/q ^ z $
for $ | q ^ z | \to {\infty } $. Closing the integration contour at infinity
one then obtains (\ref{sumrule}).


\end{document}